\documentclass[prl,twocolumn,longbibliography]{revtex4-2}
\usepackage{graphicx}  
\usepackage{dcolumn}   
\usepackage{bm}        
\usepackage{verbatim}   
\usepackage{graphicx}
\usepackage{epstopdf}
\usepackage{footnote}
\usepackage{epsfig}
\usepackage{subfigure}
\usepackage{amssymb}
\usepackage{amsmath,bm}
\usepackage{xcolor}
\usepackage{float}
\usepackage{subfigure}
\usepackage[most]{tcolorbox}
\usepackage{soul}
\usepackage{color}
\usepackage[colorlinks=true,linkcolor=blue,allcolors=blue]{hyperref}%
\usepackage[framemethod=TikZ]{mdframed}
\usepackage{lipsum}
\mdfdefinestyle{MyFrame}{%
	linecolor=blue,
	outerlinewidth=2pt,
	roundcorner=20pt,
	innertopmargin=\baselineskip,
	innerbottommargin=\baselineskip,
	innerrightmargin=20pt,
	innerleftmargin=20pt,
	backgroundcolor=gray!50!white}

\begin{document}
	
	
\headsep = 40pt
\title{Space-Time-Coupled Qubits for Enhanced Superconducting Quantum Computing}
\author{Sajjad Taravati}

\affiliation{Faculty of Engineering and Physical Sciences, University of Southampton, Southampton SO17 1BJ, UK.
}
\email{s.taravati@soton.ac.uk}
	
\begin{abstract}
The pursuit of scalable and robust quantum computing necessitates innovative approaches to overcome the inherent challenges of qubit connectivity, decoherence, and susceptibility to noise and crosstalk. Conventional monochromatic qubit coupling architectures, constrained by nearest-neighbor interactions and limited algorithmic flexibility, exacerbate these issues, hindering the realization of practical large-scale quantum processors. In this work, we introduce a paradigm leveraging a space-time-modulated cryogenic-compatible Josephson metasurface to enable polychromatic qubit coupling. This metasurface facilitates frequency-selective interactions, transforming nearest-neighbor connectivity into all-to-all qubit interactions, while significantly enhancing coherence, noise robustness, and entanglement fidelity. Our proposed approach capitalizes on the unique capabilities of space-time-modulated Josephson metasurfaces, including dynamic four-dimensional wave manipulation, nonreciprocal state transmission, and state-frequency conversion, to mediate multi-frequency qubit interactions. By isolating qubit couplings into distinct spectral channels, the cryogenic-compatible metasurface mitigates crosstalk and environmental decoherence, extending coherence times and preserving quantum state fidelity. Full-wave simulations and quantum performance analyses demonstrate a significant enhancement in the operational efficiency of a superconducting qubit array, showcasing improved connectivity, robustness, and entanglement stability. This study establishes the potential of space-time-modulated cryogenic-compatible Josephson metasurfaces as a transformative platform for next-generation quantum computing, addressing critical bottlenecks and paving the way for scalable, high-performance quantum processors.
\end{abstract}
	
\maketitle

\section{Introduction}
The advent of quantum computing promises transformative advancements across numerous fields, including cryptography, optimization, materials science, and artificial intelligence~\cite{divincenzo1995quantum,ladd2010quantum,harrow2017quantum}. Central to these advancements are qubits, the quantum counterparts of classical bits, which leverage superposition and entanglement to perform complex computations exponentially faster than classical systems~\cite{gambetta2017building,berlin2020axion,dixit2021searching,bass2024quantum}. However, current quantum computing platforms face significant challenges in scaling and reliability, primarily due to limited qubit connectivity~\cite{benjamin2001quantum,holmes2020impact,he2022control}, decoherence~\cite{martinis2005decoherence,mcdermott2009materials}, and susceptibility to noise and crosstalk~\cite{mundada2019suppression,tripathi2022suppression}. Addressing these limitations is crucial for the realization of practical and large-scale quantum computing.

Conventional quantum computing architectures predominantly rely on monochromatic operations, where qubits interact at a single frequency and primarily couple with adjacent neighbors~\cite{einstein1935can,horodecki2009quantum,ladd2010quantum,erhard2020advances}. While this nearest-neighbor connectivity simplifies control and reduces cross-interference, it imposes severe restrictions on algorithmic flexibility and computational efficiency. Non-adjacent qubits are typically isolated, requiring intricate gate operations or auxiliary qubits to facilitate long-range interactions. Moreover, the monochromatic nature of interactions exacerbates susceptibility to crosstalk and noise, further limiting coherence times and computational accuracy.

Metasurfaces, two-dimensional compact metamaterials, have revolutionized the field of electromagnetics and photonics by offering unparalleled control over the amplitude, phase, and polarization of light~\cite{taravati2020full,Taravati_NC_2021}. These engineered surfaces enable the precise manipulation of electromagnetic waves, facilitating compact and efficient solutions for a wide range of applications, including imaging, communication, and sensing.

In recent years, the advent of space-time-modulated metasurfaces has opened exciting new frontiers, introducing the concept of four-dimensional light manipulation~\cite{Taravati_Kishk_MicMag_2019,Taravati_Kishk_TAP_2019,Taravati_ACSP_2022,sisler2024electrically}. By dynamically modulating both the spatial and temporal properties of electromagnetic waves, these advanced metasurfaces provide capabilities beyond traditional static designs. They enable groundbreaking functionalities such as nonreciprocal light propagation~\cite{Taravati_PRB_SB_2017,Taravati_PRAp_2018,Taravati_Kishk_PRB_2018,Taravati_AMTech_2021,taravati2024spatiotemporal,nagulu2024synthetic}, frequency conversion~\cite{Taravati_PRB_Mixer_2018,taravati2021pure,Taravati_AMA_PRApp_2020}, and spatiotemporal beam shaping~\cite{taravati_PRApp_2019,taravati20234d}, making them indispensable tools for modern electromagnetics and photonic systems and emerging quantum technologies. Despite these advances, conventional time-modulated systems, including frequency converters and components like varactors, transistors, and diodes, are fundamentally constrained in millikelvin-temperature environments typical of superconducting quantum technologies, where their operational inefficiencies and noise generation present significant challenges.

To overcome these challenges, we propose a novel paradigm leveraging a space-time-modulated Josephson metasurface to mediate polychromatic coupling between qubits. This cryogenic-compatible dynamic metasurface introduces frequency-selective interactions, enabling all-to-all qubit connectivity while simultaneously improving coherence and robustness against environmental perturbations. By coupling adjacent and non-adjacent qubits at multiple distinct frequencies, the metasurface expands the quantum processor’s operational bandwidth, providing several key advantages as follows.

1. Enhanced connectivity and algorithm efficiency: The metasurface facilitates polychromatic interactions, transforming limited nearest-neighbor coupling into all-to-all connectivity. This enhanced connectivity enables more efficient implementation of complex quantum algorithms, reducing the number of required gate operations and their associated errors.

2. Improved coherence through frequency separation: Frequency-division multiplexing reduces crosstalk by isolating qubit interactions into distinct spectral channels, minimizing interference and preserving quantum state fidelity.

3. Coherence time improvement: By mitigating noise and environmental decoherence through tailored coupling mechanisms, the metasurface extends qubit coherence times, allowing for longer and more reliable quantum computations.

4. Robustness against crosstalk and noise: The space-time modulation of the metasurface suppresses spurious interactions, enhancing the system’s overall noise immunity and enabling more precise qubit control.

5. Improved entanglement robustness: Polychromatic coupling stabilizes entangled states by reducing the impact of decoherence and cross-interference, critical for the performance of quantum error correction and multi-qubit operations.

This paper is organized as follows: Section II discusses the limitations of monochromatic qubit coupling and outlines the theoretical framework for polychromatic interactions. Section III introduces the design and operation of the space-time-modulated Josephson metasurface. Section IV presents simulation results and performance metrics, including enhanced coherence times and entanglement fidelity. Finally, Section V concludes with an outlook on future applications and implications of this work in quantum computing.

\section{Theoretical Implications}~\label{sec:theo}
\subsection{Concept of Space-Time-Coupled Qubits}~\label{sec:concept}
\begin{figure*}
	\begin{center}
		\includegraphics[width=0.8\linewidth]{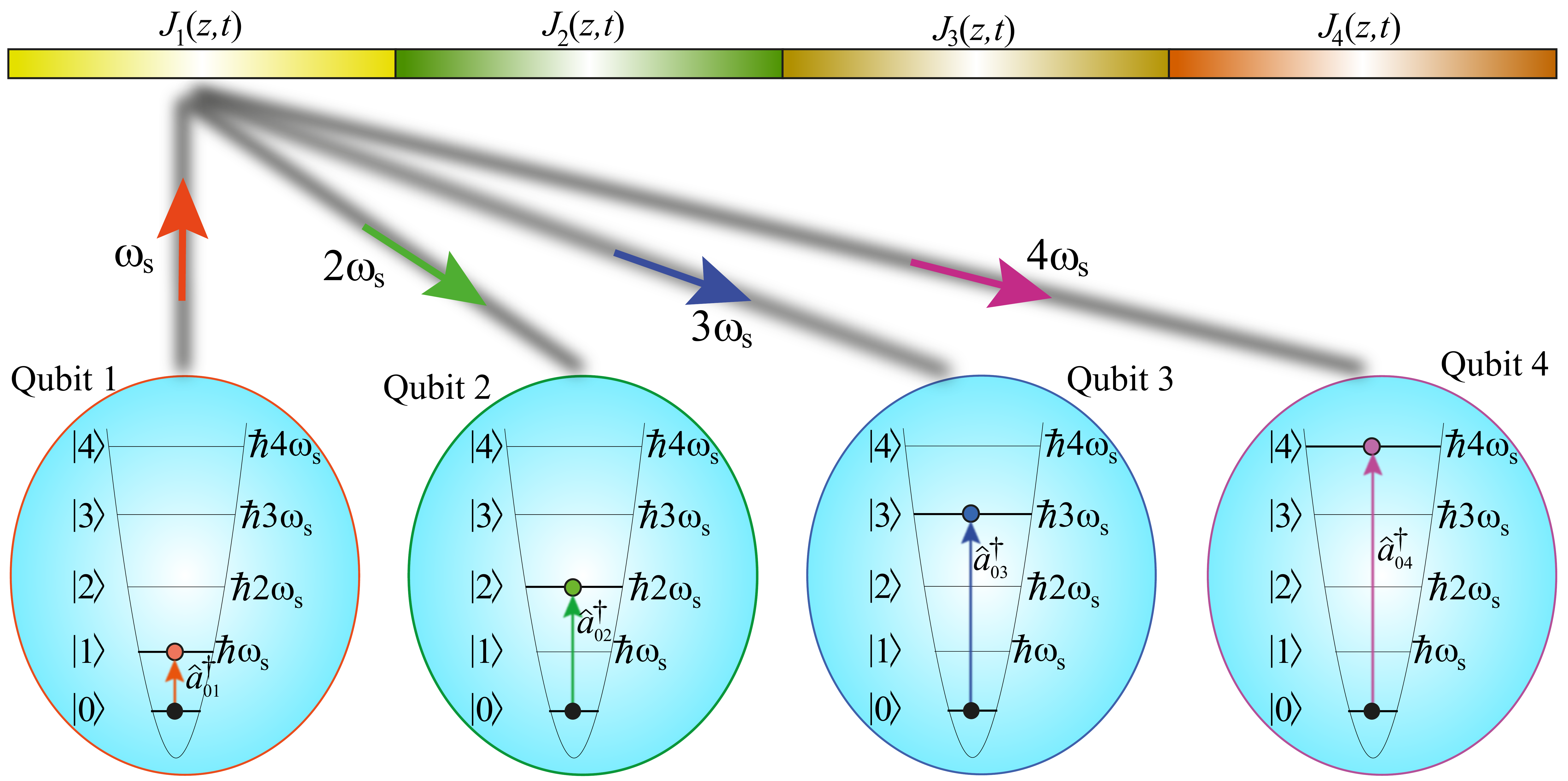}  
		\caption{Schematic representation of space-time-modulated coupling between superconducting qubits, demonstrating energy transfer and selective excitation (e.g., Qubit 1 at state $\hbar \omega_\text{s}$ exciting Qubits 2–4 at distinct quantum states $\hbar 2\omega_\text{s}$, $\hbar 3\omega_\text{s}$ and $\hbar 4\omega_\text{s}$). The process is mediated by a reflective, cryogenic-compatible superconducting metasurface with space-time periodicity, enabling multi-frequency control of qubit interactions and selective state transitions to enhance quantum computing efficiency.}
		\label{Fig:energylevels}
	\end{center}
\end{figure*}

Figure~\ref{Fig:energylevels} illustrates a schematic representation of energy transfer and state-selective excitation among four frequency-distinct qubits mediated by a reflective space-time-periodic superconducting cryogenic-compatible metasurface. The system comprises four qubits with discrete quantum energy states ($h\omega_\text{s}$, $2h\omega_\text{s}$, $3h\omega_\text{s}$, and $4h\omega_\text{s}$), corresponding to distinct operational frequencies. The metasurface, composed of four regions with space-time-modulated critical current densities ($J_1(z,t)$, $J_2(z,t)$, $J_3(z,t)$, and $J_4(z,t)$), is designed to selectively interact with each qubit and mediate controlled quantum state transitions. A photon emitted by Qubit 1 at frequency $\omega_\text{s}$ is reflected by the metasurface, which employs space-time modulation to redistribute the photon's energy into three distinct frequencies ($2\omega_\text{s}$, $3\omega_\text{s}$, and $4\omega_\text{s}$) while imparting specific angular momentum components. These frequency-shifted components are directed to Qubits 2, 3, and 4, resonantly exciting them from their respective ground states to higher energy states: $2h\omega_\text{s}$, $3h\omega_\text{s}$, and $4h\omega_\text{s}$.  The space-time modulation facilitates controlled quantum state conversion and enables nonreciprocal, frequency-selective coupling among both the adjacent and non-adjacent qubits. This interaction enhances qubit coherence, minimizes crosstalk, and supports high-fidelity state transfer, demonstrating the metasurface's potential to mediate multi-frequency qubit entanglement and scalable quantum computation.  

The reflective metasurface acts as a precise frequency filter, ensuring minimal energy leakage and suppressing back-reflections. This selective control reduces decoherence and enhances the fidelity of quantum state transitions. The metasurface’s tailored design minimizes environmental noise and thermal perturbations, a critical factor for maintaining coherence in quantum systems operating at cryogenic temperatures. By leveraging space-time modulation, the system introduces directional energy flow, enabling unidirectional coupling between qubits. This nonreciprocity ensures that state transitions occur in a controlled and predictable manner, a crucial requirement for robust quantum gates and entanglement protocols. The metasurface’s modular design supports interactions among multiple qubits operating at distinct frequencies. This scalability is essential for constructing large-scale quantum processors where qubits may have different operational characteristics. State-frequency conversion allows heterogeneous qubit architectures (e.g., superconducting and photonic qubits) to interact seamlessly, broadening the applicability of hybrid quantum systems. 

\subsection{Limitation of Conventional Qubit Arrays}
\begin{figure}
	\begin{center}
		\includegraphics[width=1\linewidth]{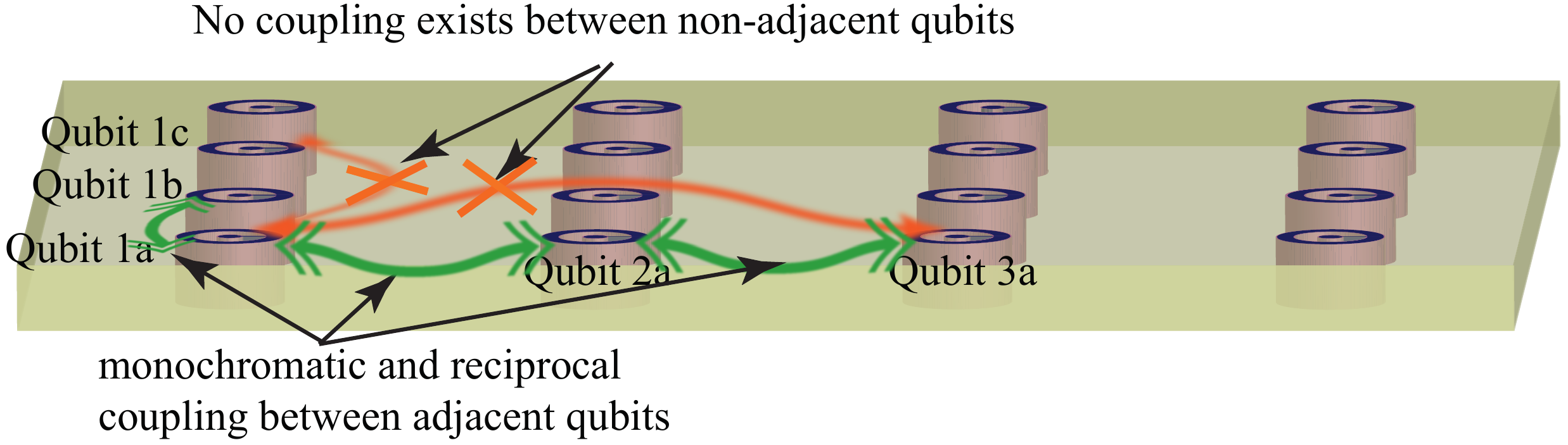}  
		\caption{Illustration of the restricted functionality in conventional quantum computing qubit arrays, where coupling is limited to \textit{monochromatic} interactions between \textit{adjacent} qubits, preventing direct non-adjacent interactions and rendering the architecture incompatible with arrays of distinct-frequency qubits.}
		\label{Fig:sch_a}
	\end{center}
\end{figure}

Figure~\ref{Fig:sch_a} depicts a traditional quantum computing qubit arrays operate at single frequencies, limiting direct interactions to systems with matched resonance. The metasurface overcomes this limitation by converting a single input frequency into multiple, targeted frequencies, effectively enabling polychromatic qubit coupling. This coupling mechanism supports complex quantum operations such as multi-qubit entanglement, quantum error correction, and advanced quantum algorithms. The metasurface can mediate entanglement between qubits across different frequencies, an essential feature for distributed quantum networks and algorithms requiring frequency diversity. The ability to transition qubits between specific energy states through controlled interactions enhances quantum gate operations and state preparation techniques. The metasurface’s nonreciprocal and frequency-selective properties shield the qubits from unwanted noise, improving the robustness of quantum computations.  

In conventional superconducting quantum computing platforms, the interaction (coupling) between qubits is typically mediated through a shared coupling mechanism such as a resonator or a direct coupling element like a capacitive or inductive link. These systems often operate at a single frequency, which we refer to as monochromatic interaction. In such setups, only adjacent qubits couple strongly, while non-adjacent qubits experience negligible coupling due to the rapid spatial decay of the interaction strength. This limitation arises because the interaction is usually designed to suppress crosstalk and ensure precise control of individual qubit operations. Consider a 4×4 superconducting qubit array with qubits indexed by their positions $(i, j)$, where $i$ and $j$ represent the row and column indices. The Hamiltonian for this system reads
\begin{subequations}
\begin{equation}\label{eqa:H1}
		\begin{split}
H =& \sum_{(i,j)} \frac{\omega_{ij}}{2} \sigma_z^{(i,j)} \\&+ \sum_{\langle (i,j), (k,l) \rangle} g_{ij,kl} \left( \sigma_+^{(i,j)} \sigma_-^{(k,l)} + \sigma_-^{(i,j)} \sigma_+^{(k,l)} \right),
	\end{split}
\end{equation}
where $\omega_{ij}$ is the resonance frequency of the qubit at $(i, j)$, $g_{ij,kl}$ is the coupling strength between adjacent qubits $(i, j)$ and $(k, l)$, $\sigma_z^{(i,j)}$ is the Pauli-$z$ operator for the qubit at $(i, j)$, $\sigma_+^{(i,j)}$ and $\sigma_-^{(i,j)}$ are the raising and lowering operators for the qubit at $(i, j)$, respectively, and $\langle (i,j), (k,l) \rangle$ denotes summation over adjacent qubits only. The assumptions for conventional monochromatic coupling between qubits are that all qubits are designed to operate at the same or nearly the same resonance frequency $\omega_q$, assuming $\omega_{ij} \approx \omega_q$, and the coupling strength $g_{ij,kl}$ is non-zero only for adjacent qubits, and it decays rapidly for non-adjacent qubits. For adjacent qubits $(i, j)$ and $(k, l)$, the interaction term simplifies to
\begin{equation}\label{eqa:H2}
H_{\text{int}} = g_{ij,kl} \left( \sigma_+^{(i,j)} \sigma_-^{(k,l)} + \sigma_-^{(i,j)} \sigma_+^{(k,l)} \right).
\end{equation}

For a single pair of adjacent qubits, the time evolution under this Hamiltonian leads to the standard Rabi oscillation between their states. For non-adjacent qubits, the coupling strength $g_{ij,kl}$ is negligible due to spatial separation and the design of the coupling network, as
\begin{equation}\label{eqa:g}
	\begin{split}
g_{ij,kl} \propto \frac{1}{d^n},
	\end{split}
\end{equation}
\end{subequations}
where $d$ is the spatial distance between the qubits, and $n \geq 3$ for typical superconducting qubit systems, ensuring that $g_{ij,kl} \approx 0$ for non-adjacent pairs.

\subsection{Long-Range and Polychromatic Entanglement Through Space-Time-Coupled Qubit Array}

\begin{figure*}
	\begin{center}
		\includegraphics[width=0.7\linewidth]{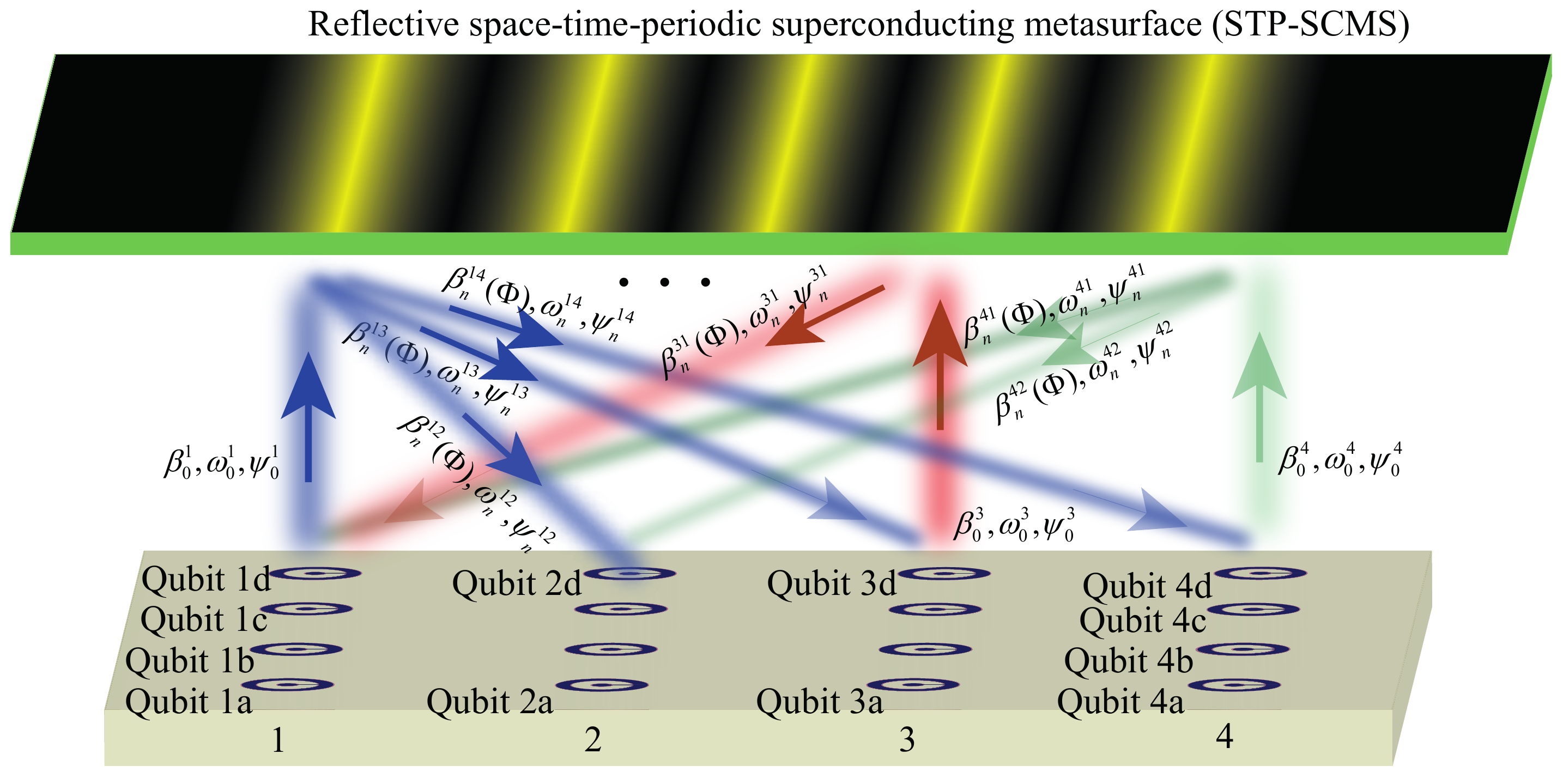} 
		\caption{Illustration of long-range, polychromatic entanglement within a qubit array enabled by a space-time-modulated cryogenic-compatible superconducting Josephson metasurface, facilitating mediated coupling between both adjacent and non-adjacent qubits across multiple frequencies.}
		\label{Fig:sch_b}
	\end{center}
\end{figure*}

We now propose the integration of a space-time-modulated cryogenic-compatible Josephson metasurface atop the superconducting qubit array, as depicted in Fig.~\ref{Fig:sch_b}. This metasurface exhibits a space-time-periodic modulation characterized by a nonlinear and periodic critical current density, expressed as
\begin{equation}\label{eqa:J}
	J(z,t) = f_\text{nl,per}(\kappa_\text{s} z, \omega_\text{s} t),
\end{equation}
where $f_\text{nl,per}$(.) denotes an arbitrary nonlinear and periodic function of space and time, with $\kappa_\text{s}$ and $\omega_\text{s}$ representing the spatial and temporal modulation frequencies of the superconducting space-time modulation, respectively. The introduction of this metasurface enables multi-frequency interactions, which are instrumental in facilitating long-range entanglement and coupling not only between adjacent qubits but also between non-adjacent qubits. This capability significantly enhances the connectivity and coherence properties of the qubit array, offering new avenues for scalable quantum information processing. The Hamiltonian in this case extends to:
\begin{equation}\label{eqa:H}
	\begin{split}
		H = \sum_{(i,j)} 	&\frac{\omega_{ij}}{2} \sigma_z^{(i,j)} + \sum_{(i,j) \neq (k,l)} \sum_m g_{ij,kl}^{(m)} \\& \cdot e^{i \Delta \omega_{ij,kl}^{(m)} t} \left( \sigma_+^{(i,j)} \sigma_-^{(k,l)} + \sigma_-^{(i,j)} \sigma_+^{(k,l)} \right),
	\end{split}
\end{equation}
where $g_{ij,kl}^{(m)}$ is the coupling strength at the $m$-th frequency and $\Delta \omega_{ij,kl}^{(m)}$ is the frequency detuning between qubits $(i,j)$ and $(k,l)$ mediated by the metasurface. Such a polychromatic interaction offers two key advantages as follows.  
1. Non-adjacent coupling: This allows for non-zero $g_{ij,kl}^{(m)}$ between qubits separated by larger distances.  
2. Frequency-multiplexed interaction: Coupling occurs at multiple frequencies, enabling enhanced connectivity.  

In the monochromatic case ($\Delta \omega = 0 $), $g_{ij,kl} \neq 0$ only for adjacent qubits. However, in the polychromatic case, coupling is mediated by multiple harmonics, where $g_{ij,kl}^{(m)}$ facilitates interactions between non-adjacent qubits. The introduction of enhanced interaction and polychromatic coupling/entanglement through a space-time-modulated Josephson metasurface has profound implications for improving the performance of quantum computing systems. Below, we detail key aspects of performance improvement and provide corresponding mathematical formulations.

\subsection{Enhanced Quantum Connectivity and Coherence}

\subsubsection{Reduction in Gate Depth Due to Enhanced Connectivity}

In a standard superconducting quantum processor only adjacent qubits couple strongly ($g_{ij,kl} \neq 0$ for adjacent pairs). In addition, quantum gates between non-adjacent qubits require multi-step swaps, increasing gate depth and computation time. The reflective space-time superconducting metasurface facilitates direct interaction between non-adjacent qubits, effectively reducing the gate depth for multi-qubit operations. If the average gate depth for a multi-qubit operation in the monochromatic case is $D_m$, the gate depth in the polychromatic case $D_p$ satisfies
\begin{subequations}
\begin{equation}\label{eqa:D}
D_p = \frac{D_m}{\langle C_{\text{polychromatic}} \rangle},
\end{equation}
where $\langle C_{\text{polychromatic}} \rangle$ represents the effective increase in connectivity facilitated by the metasurface. For a 4×4 array
\begin{equation}\label{eqa:C}
\langle C_{\text{polychromatic}} \rangle = \frac{\text{Directly coupled pairs (polychromatic)}}{\text{Directly coupled pairs (monochromatic)}}.
\end{equation}
\end{subequations}

This enhanced connectivity enables faster execution of quantum algorithms, particularly those relying on long-range entanglement, such as Grover's search or Shor's algorithm.

\subsubsection{Coherence Time Improvement}
The coherence time \( T_2 \) is a critical parameter in quantum computing, as it determines the duration over which a qubit can maintain its quantum state with minimal decoherence. In traditional (monochromatic) qubit systems, \( T_2 \) is often limited by factors such as environmental noise, crosstalk between qubits, and imperfections in control fields. The introduction of a reflective space-time-periodic metasurface enables polychromatic operation, which introduces a frequency separation \( \Delta f \) between interacting qubits. This frequency separation mitigates crosstalk, reduces overlap in spectral components, and effectively enhances the coherence time of the system. The modified coherence time in the polychromatic case is given by  
\begin{equation}\label{eqa:ST}
	T_2' = T_2 \left(1 + 2\pi\frac{\Delta f}{\gamma}\right),
\end{equation}
where \( \Delta f \) is the frequency separation introduced by the metasurface, which increases the distinguishability of qubits operating at different frequencies, and \( \gamma \) is the decoherence rate due to crosstalk and other noise sources. The metasurface creates distinct operational frequencies for qubits, effectively isolating their quantum states in the frequency domain. This reduces spectral overlap, which is a significant source of qubit interaction errors and decoherence in densely packed quantum systems. The term \( \frac{\Delta f}{\gamma} \) reflects how increasing the separation between frequencies proportionally enhances the coherence time. Crosstalk arises when qubits inadvertently interact due to overlapping frequency components or shared control fields. By introducing a separation \( \Delta f \), the metasurface reduces these unwanted interactions, lowering the effective decoherence rate \( \gamma \). Consequently, the system becomes more robust against external perturbations, allowing for longer coherence times.

An increased coherence time \( T_2' \) directly enhances the fidelity of quantum gate operations, as qubits remain in well-defined quantum states for longer durations. This improvement is particularly beneficial for complex algorithms and error correction protocols, which require sustained coherence across multiple computational steps. As quantum systems scale to include more qubits, crosstalk and decoherence become increasingly problematic. The polychromatic approach provides a scalable solution by enabling qubits to operate at distinct frequencies, reducing inter-qubit interference in larger systems. This scalability is essential for the development of practical, high-performance quantum computers. The enhancement in \( T_2' \) is contingent on the design parameters of the metasurface. Larger \( \Delta f \) values can be achieved by optimizing the space-time modulation parameters, such as modulation frequency and spatial periodicity. However, these design choices must balance other factors, such as the bandwidth and angular dispersion, to ensure efficient energy transfer and qubit excitation.

This enhancement in coherence time contributes to several quantum computing advancements as follows. Longer coherence times lead to fewer errors during gate operations, improving the overall computation fidelity. Furthermore, extended coherence times reduce the overhead required for error correction, allowing more computational resources to be directed toward algorithmic tasks. Additionally, the frequency-selective approach facilitates interactions between heterogeneous qubits, enabling hybrid systems (e.g., superconducting and photonic qubits) with improved coherence properties. The introduction of a frequency separation $\Delta f$ between interacting qubits reduces the spectral overlap between their noise spectra. This reduction in spectral overlap leads to a decrease in the effective decoherence rate $\gamma$, resulting in an increase in the coherence time. Crosstalk between qubits is a significant source of decoherence in quantum computing systems. By introducing a frequency separation $\Delta f$, the metasurface suppresses crosstalk between qubits, reducing the effective decoherence rate and increasing the coherence time. The metasurface creates distinct operational frequencies for qubits, effectively isolating their quantum states in the frequency domain. This isolation reduces the impact of external noise sources and crosstalk, leading to an increase in the coherence time. The increase in coherence time enabled by the metasurface is crucial for the scalability of quantum computing systems. As the number of qubits increases, the coherence time typically decreases due to increased crosstalk and spectral overlap. The metasurface mitigates this effect, enabling the development of larger-scale quantum computing systems with improved coherence times.

\subsubsection{Robustness Against Crosstalk and Noise}
Monochromatic systems are prone to interference from thermal noise and material defects, and suffer from crosstalk between adjacent qubits due to shared resonators. The metasurface spatially and temporally modulates the interaction, dynamically isolating qubits from common-mode noise. 	The metasurface introduces a structured interaction by modulating the spatial and temporal properties of the qubit-resonator environment. This modulation redistributes the noise spectral density \( S_0(\omega) \), effectively suppressing unwanted contributions within a specified frequency band around the qubit's operational frequency \( \omega_q \). The effective noise spectral density $S_\text{eff}(\omega)$ decreases as:
\begin{equation}\label{eqa:S}
S_\text{eff}(\omega) = S_0(\omega) \cdot \exp\left(-\frac{\Delta \omega^2}{2\sigma^2}\right),
\end{equation}
where $S_0(\omega)$ is initial noise spectral density and $\sigma$ is the bandwidth of noise suppression around $\omega_q$. This results in a significant reduction in dephasing noise, improving both $T_1$ and $T_2$. 	This exponential attenuation ensures that noise contributions at frequencies distant from \( \omega_q \) are significantly diminished, reducing their impact on qubit performance. The suppression mechanism is not static; it dynamically adapts to changes in qubit frequencies, ensuring consistent protection even in systems with varying operational parameters.

Monochromatic quantum systems, while foundational to early quantum computing, exhibit inherent vulnerabilities that limit their scalability and performance. These systems are particularly susceptible to two critical issues: as follows. Thermal noise introduces random fluctuations in the system, degrading qubit coherence by inducing energy transitions or phase shifts. Material defects, such as impurities or structural inconsistencies in superconducting circuits, exacerbate decoherence by contributing additional, often unpredictable, noise sources. Crosstalk arises from unintended interactions between qubits sharing resonators or coupling elements. This leads to correlated errors, reducing the fidelity of individual quantum gates and overall computation reliability. Crosstalk, a major impediment to scalable quantum systems, is addressed through the metasurface's ability to create frequency isolation between qubits. By modulating interactions at distinct frequencies for each qubit, the metasurface effectively decouples adjacent qubits, minimizing their mutual influence. This isolation reduces correlated errors and enhances gate fidelity, even in densely packed qubit arrays.  

The metasurface's noise suppression and crosstalk mitigation mechanisms translate to tangible benefits in quantum computing systems as follows. Firstly, by attenuating \( S_\text{eff}(\omega) \), the metasurface minimizes phase errors, directly improving coherence times \( T_1 \) (energy relaxation) and \( T_2 \) (dephasing). Secondly, isolated qubit interactions ensure that quantum gates operate with minimal error, reducing the overhead for error correction and boosting computational accuracy. Thirdly, robust noise suppression and crosstalk mitigation enable closer qubit packing and support for larger qubit arrays, paving the way for scalable quantum processors. Finally, the metasurface can be reconfigured to accommodate changes in qubit frequencies or operational requirements, making it versatile for evolving quantum computing architectures.

\subsubsection{Improved Entanglement Robustness}

Qubit Fidelity is a measure of the accuracy and reliability of a qubit or a quantum operation. In quantum computing, fidelity quantifies how closely the actual state or operation aligns with the intended or ideal state or operation. High fidelity is crucial for the performance of quantum algorithms, as errors and noise can quickly accumulate and render computations meaningless. In monochromatic systems, entanglement fidelity is limited by unwanted interactions due to frequency crowding, and decoherence from shared noise sources. Fidelity, in the context of quantum mechanics, is a measure of how close two quantum states are. It can be defined for two quantum states $ \rho $ and $ \sigma $ as
\begin{subequations} 
\begin{equation}\label{eqa:F1}
F(\rho, \sigma) = \left( \text{Tr} \left( \sqrt{ \sqrt{\rho} \sigma \sqrt{\rho} } \right) \right)^2.
\end{equation}

For coherence time, fidelity can be related to the decay of coherence over time, which typically occurs due to environmental interactions, such as in a quantum system undergoing decoherence. In such cases, the fidelity at time $ t $, given the initial state $ \rho(0) $ and a noisy, decohered state $ \rho(t) $, is often modeled as $F(t) = \exp\left(-t/T_2\right)$~\cite{liu2005fidelity,wang2005uniform,silvestrov2003hypersensitivity}, where $ T_2 $ is the coherence time of the system and $ t $ is the time elapsed. The above formula assumes that the system undergoes a simple exponential decay of coherence (which is typical in many systems such as qubits), where the fidelity decays over time as a result of the loss of coherence. For more complex cases, the fidelity can also be influenced by factors like the noise model, but the dependence on $ T_2 $ is often central in many quantum systems.

The metasurface’s polychromatic coupling architecture enhances entanglement robustness by mediating multi-frequency interactions. This approach suppresses noise-induced decoherence while enabling direct entanglement between non-adjacent qubits, thereby expanding the Hilbert space accessible for complex entangled states. Under Markovian decoherence~\cite{nielsen2010quantum,preskill2018quantum}—a regime applicable to superconducting qubits dominated by short-correlated noise—the entanglement fidelity \( F_{\text{entangled}}' \) in the polychromatic case reads
\begin{equation}\label{eqa:F}
	F_{\text{entangled}}' = F_{\text{entangled}} \cdot \exp\left(-\frac{t}{T_2'}\right),
\end{equation}  
\end{subequations} 
which is the enhanced coherence time enabled by frequency separation \( \Delta f \). The exponential fidelity decay slows proportionally to \( T_2' \), ensuring \( F_{\text{entangled}}' > F_{\text{entangled}} \) for identical noise conditions. This fidelity improvement arises from the metasurface’s spectral isolation of qubit interactions, which reduces overlap with broadband noise and suppresses correlated errors in multi-qubit operations.

\section{Results}
We consider a periodic space-time-varying Josephson junction array, where the inductance of the dynamic superconducting metasurface is expressed as~\cite{taravati2024efficient,taravati2025light}
\begin{equation}\label{eqa:L}
	\begin{split}
		L_\text{s}(I,z,t)
		= \dfrac{\Phi_0}{ 2\pi I_0}  \sec\left(\widetilde{\Phi}_\text{dc}+ \widetilde{\Phi}_\text{rf} \sin[\kappa_\text{s} z-\omega_\text{s} t+\phi]\right),
	\end{split}
\end{equation}
where $\widetilde{\Phi}_\text{dc}= 2\pi \Phi_\text{dc}/ \Phi_0$ and $\widetilde{\Phi}_\text{rf}= 2\pi \Phi_\text{rf}/ \Phi_0$, $\Phi_0$ is the magnetic flux quantum. Considering the incident field $\mathbf{H}_\text{I} (x,z,t)= \mathbf{\hat{y}} H_0 e^{j \omega _0 t}\cdot e^{-j\left[k_0\sin(\theta_\text{i}) x +k_0 \cos(\theta_\text{i}) z \right]}$, the reflected fields reads
\begin{equation}
	\mathbf{H}_\text{R} (x,z,t)= \mathbf{\hat{y}} \sum\limits_{n =  - \infty }^\infty  R_{n} e^{-j \left[ k_{0} \sin(\theta_{\text{i}}) x -k_{0n} \cos(\theta_n^\text{R}) z  -\omega_n t\right] } ,
	\label{eqa:A-E_refl_forw}
\end{equation}
\noindent where $\omega_n=\omega_0+n \omega_\text{s}$, $\theta_\text{i}$ is defined as the angle between the incident wave and the metasurface boundary, and $\theta_n^\text{R}$ is defined as the angle between the $n$th reflected space-time harmonic and the metasurface boundary. We apply continuity of the tangential components of the electromagnetic fields at $z=0$ and $z=d$ to find the unknown field amplitudes $R_{n}$~\cite{taravati2025light}. 

\begin{figure}
	\begin{center}
		\includegraphics[width=1\linewidth]{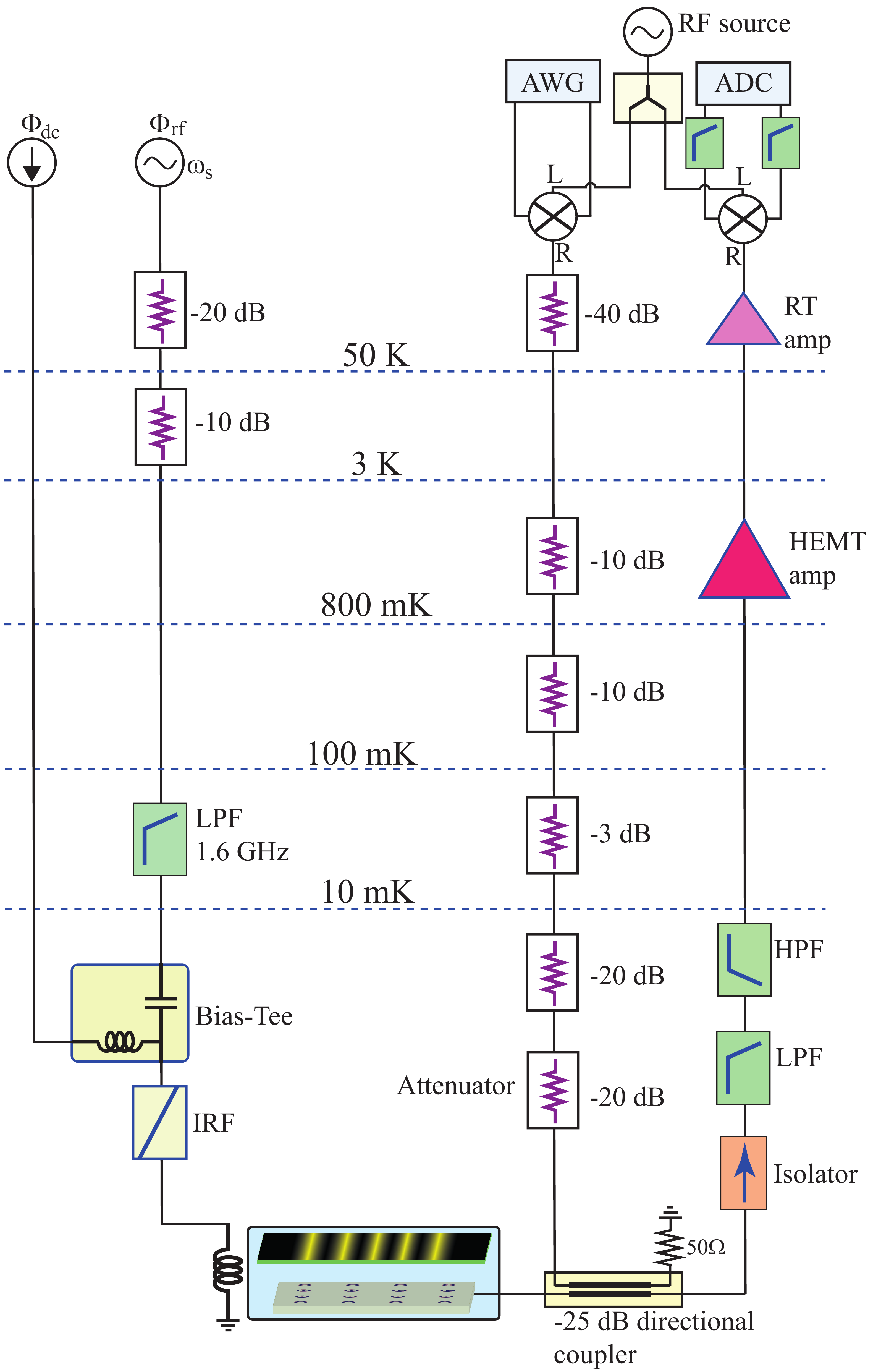}  
		\caption{Experimental setup for quantum circuit characterization. The setup includes a multilayered cryogenic chain with attenuators, filters, and thermal isolation for noise suppression. Signal amplification is performed using a cryogenic HEMT amplifier at 3 K, followed by a room-temperature (RT) amplifier. A bias tee provides low-frequency flux bias ($\Phi_{\text{dc}}$) and RF drive ($\Phi_{\text{rf}}$) to the quantum circuit, while the output signal is filtered, amplified, mixed, and digitized at room temperature.}
		\label{Fig:meas}
	\end{center}
\end{figure}

To characterize the quantum circuit and extract the output signals, a multilayered cryogenic and amplification setup is employed, as illustrated in Fig.~\ref{Fig:meas}. The quantum circuit, located at the base temperature of 10 mK within a dilution refrigerator, is shielded from noise and thermal interference through a series of carefully designed components. A -25 dB directional coupler extracts the readout signal from the quantum circuit, which is then isolated using a cryogenic isolator to prevent back-reflections that could perturb the circuit's coherence. The extracted signal is filtered by a high-pass filter (HPF) and a low-pass filter (LPF) to ensure noise suppression and bandwidth limitation. Attenuators are strategically placed across the temperature stages (10 mK to 50 K) to suppress spurious thermal noise contributions and improve impedance matching between the stages. The infrared filter (IRF) in the superconducting qubit measurement setup is crucial for suppressing infrared (IR) radiation that can propagate through wiring and degrade qubit performance. IR radiation introduces thermal noise, which can cause decoherence or unintended qubit excitations, and can break Cooper pairs in the superconducting circuit, generating quasiparticles that reduce coherence times. Positioned after the bias tee, the IRF ensures low-frequency control and readout signals pass through while blocking high-energy photons, maintaining a noise-free and isolated environment essential for qubit operation and accurate measurements. At the 3 K stage, a cryogenic High Electron Mobility Transistor HEMT amplifier provides the first stage of amplification with a low noise figure to boost the weak signal from the qubit while preserving its integrity. The amplified signal is then routed to the room-temperature (RT) stage for further amplification by an RT amplifier. At this stage, the signal is mixed with a local oscillator signal for down-conversion into an intermediate frequency (IF) band, which is digitized by an analog-to-digital converter (ADC). For driving and controlling the quantum circuit, two input lines are used. A low-frequency flux bias ($\Phi_{\text{dc}}$) is applied via a bias tee, while a microwave drive signal ($\Phi_{\text{rf}}$) is delivered at the qubit's operating frequency. A local oscillator (LO) source and mixer are used to downconvert the high-frequency qubit signal to an intermediate frequency (IF) for digitization by the ADC. An arbitrary waveform generator (AWG) generates pulse sequences for qubit control and measurement. Thermal shielding and isolation at each cryogenic stage minimize decoherence and thermal noise, ensuring reliable operation and accurate readout of the quantum circuit. This setup enables precise characterization and control of the quantum device, facilitating robust signal extraction and analysis.

\begin{figure*}
	\begin{center}
		\subfigure[]{\label{Fig:resa}
			\includegraphics[width=0.38\linewidth]{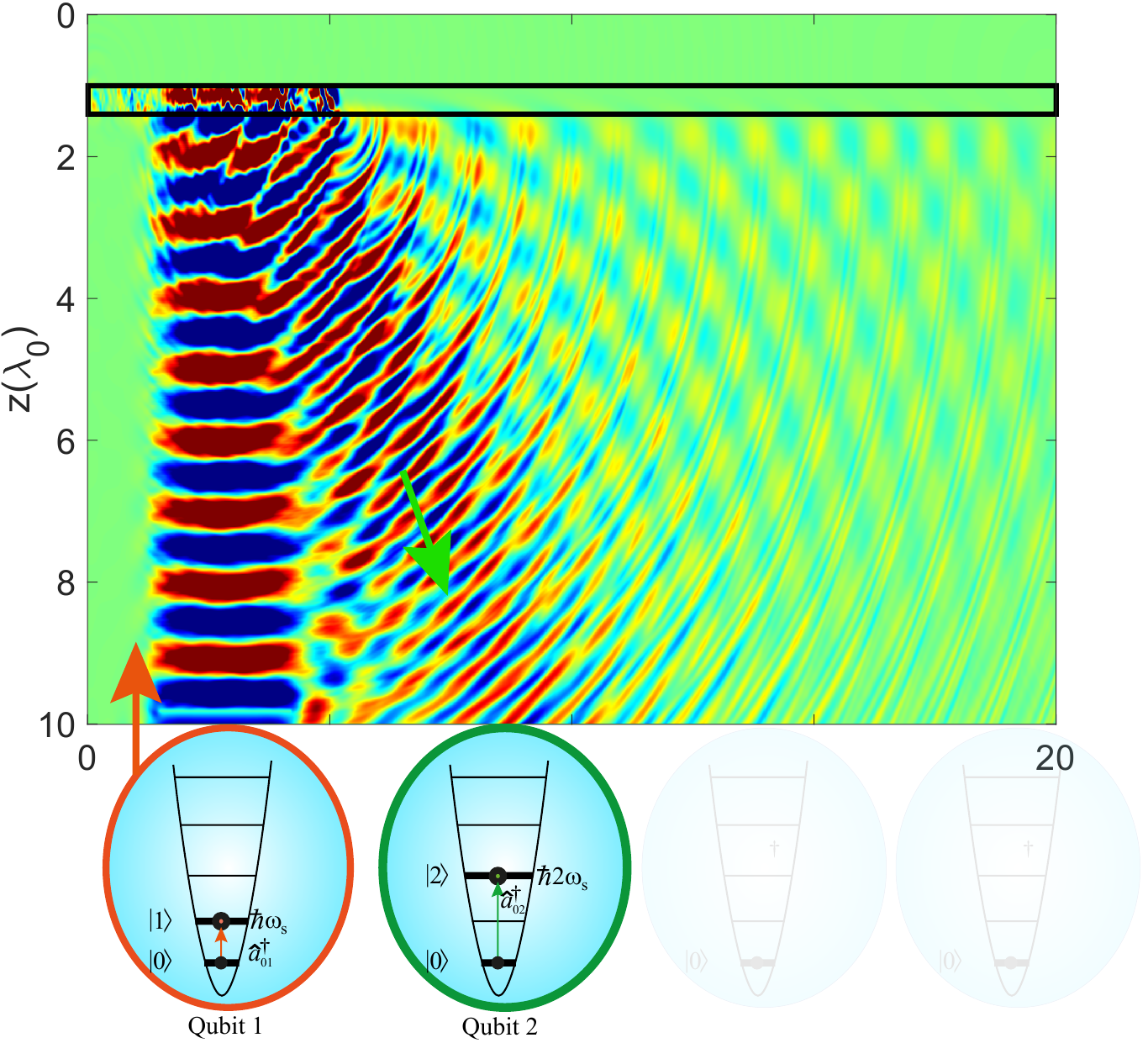}  }
		\subfigure[]{\label{Fig:resb}
			\includegraphics[width=0.38\linewidth]{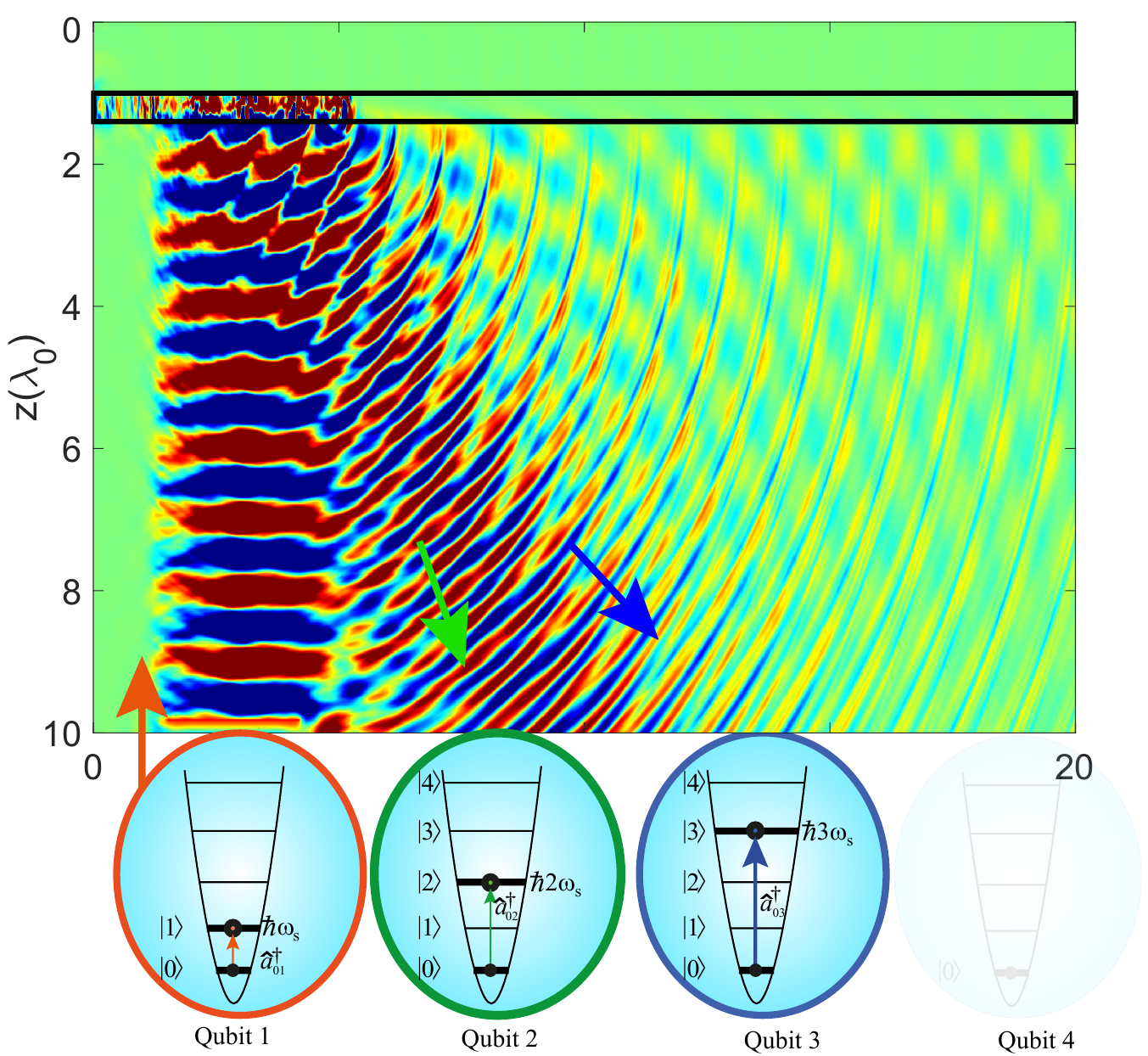}  }
		\subfigure[]{\label{Fig:resc}
			\includegraphics[width=0.38\linewidth]{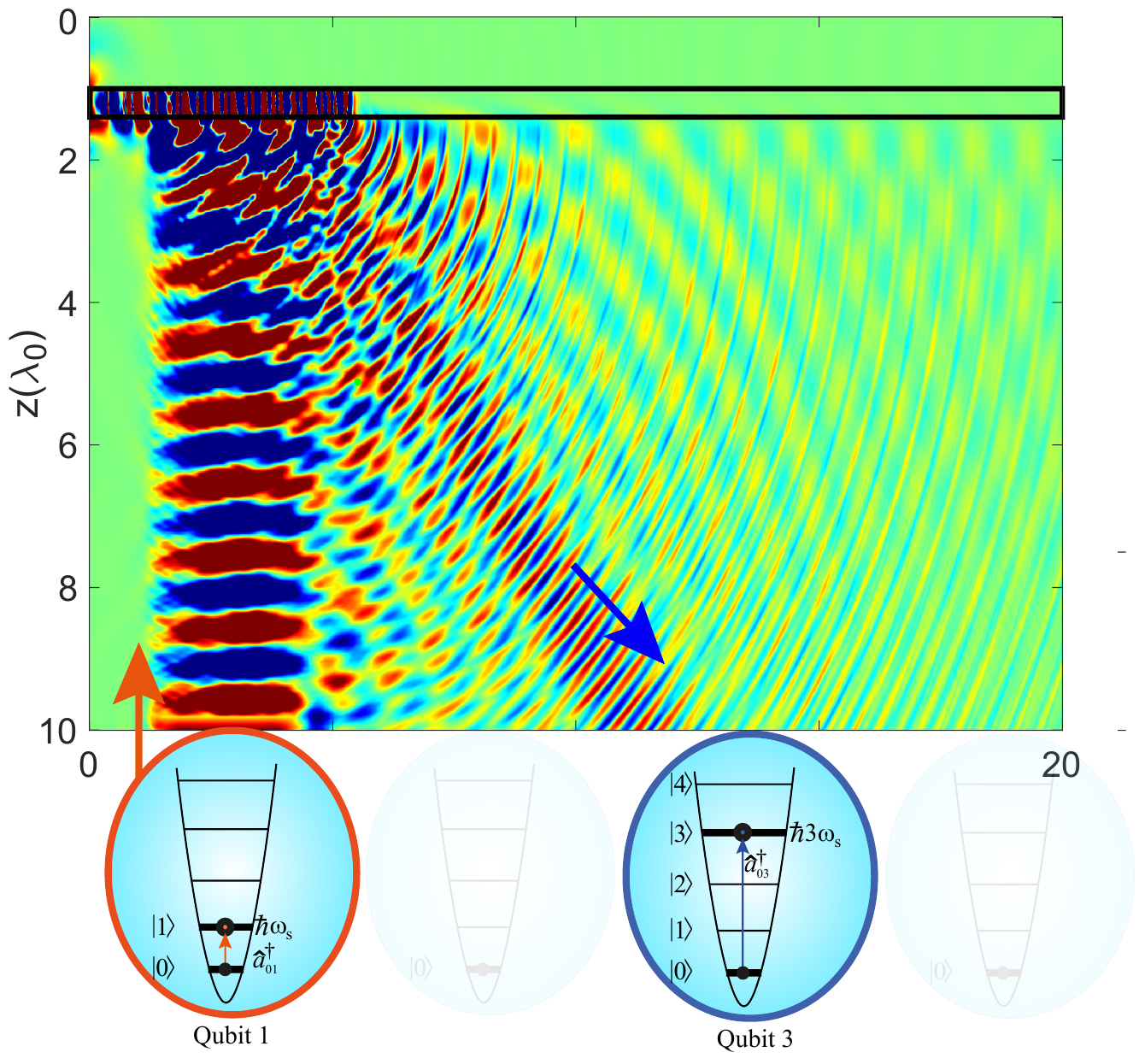}  }
		\subfigure[]{\label{Fig:resd}
			\includegraphics[width=0.38\linewidth]{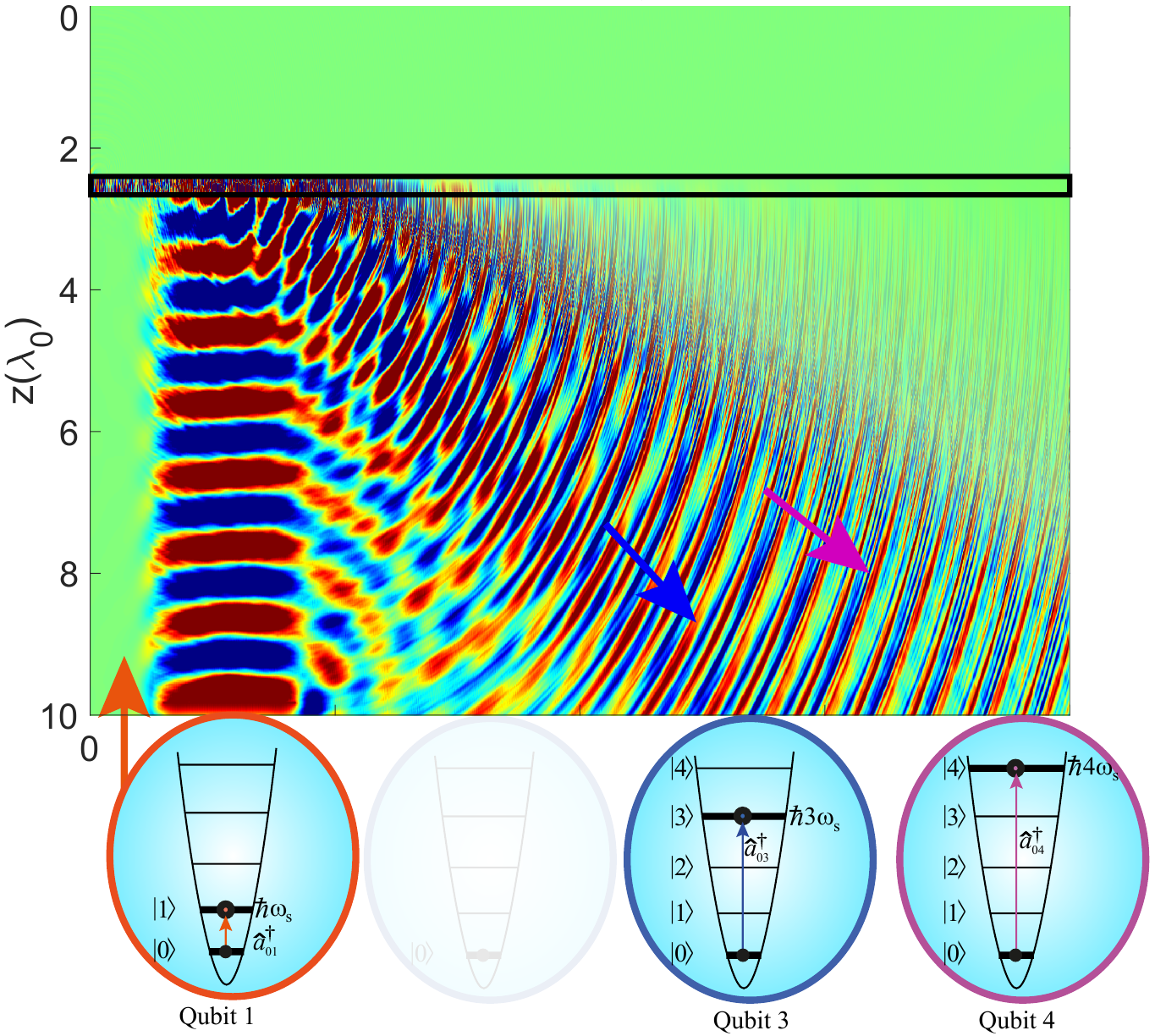}  }
		\subfigure[]{\label{Fig:rese}
			\includegraphics[width=0.38\linewidth]{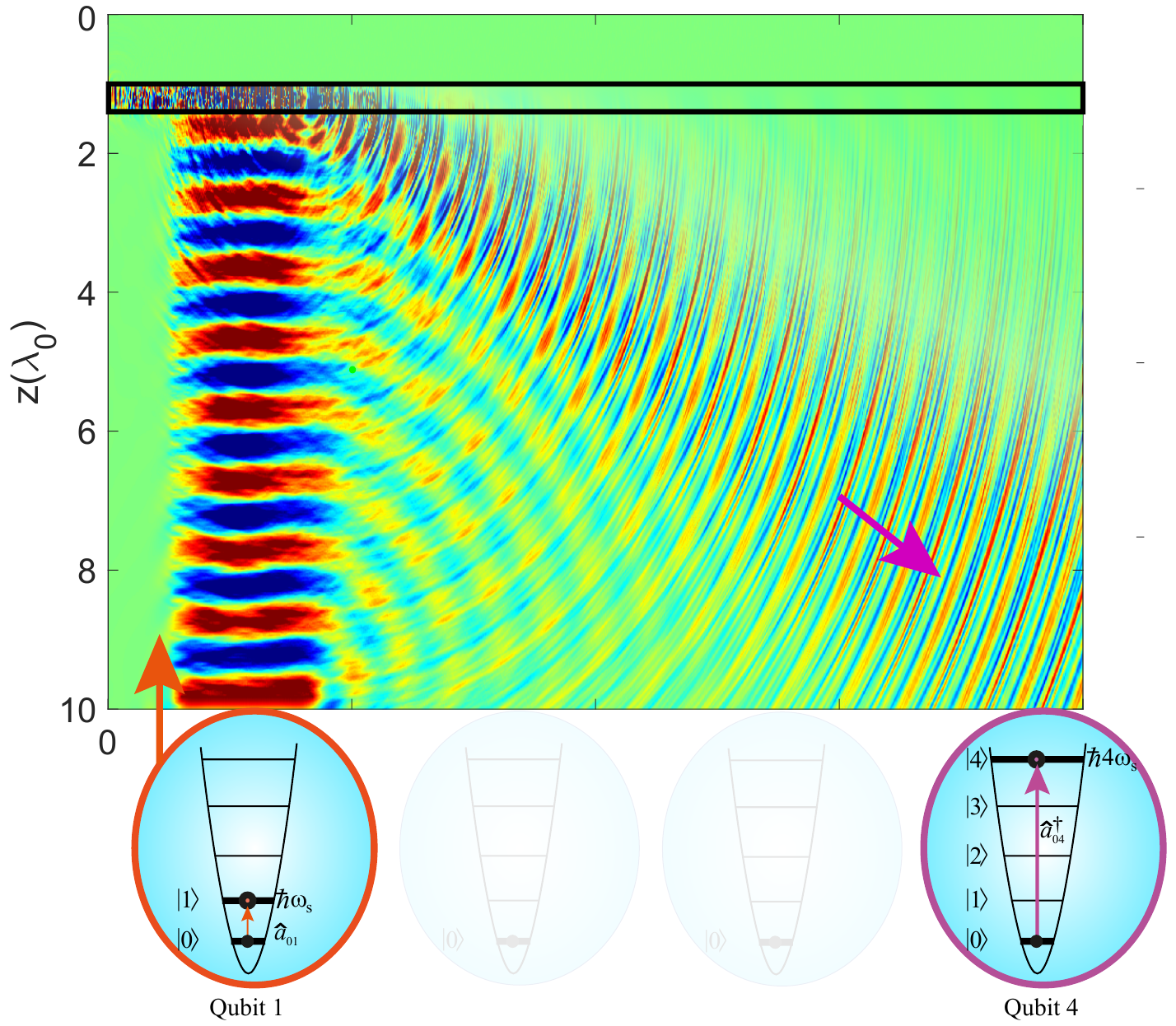}  }
		\subfigure[]{\label{Fig:resf}
			\includegraphics[width=0.38\linewidth]{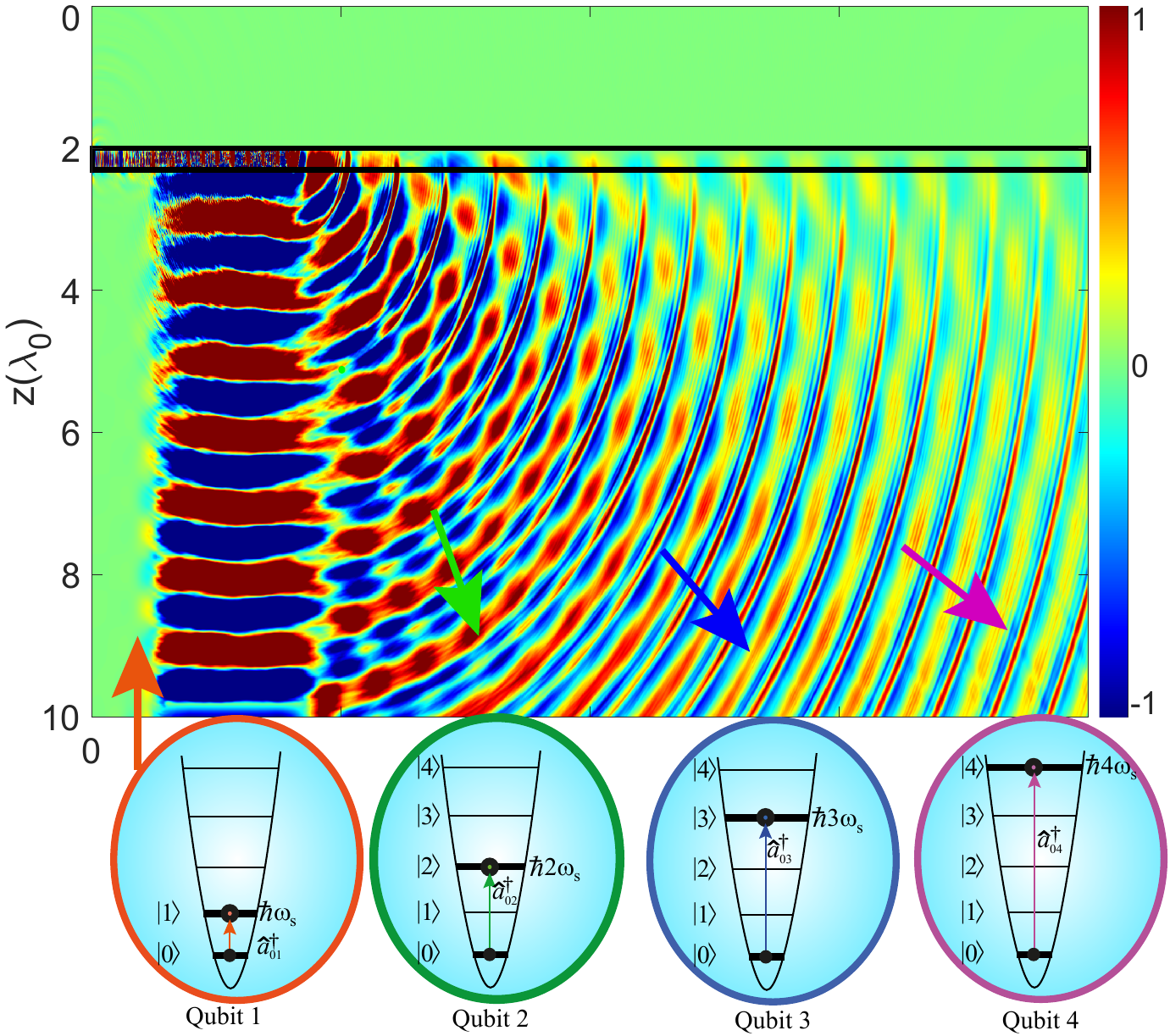}  }
		\caption{Selective space-time coupling and frequency-state transitions of superconducting qubits mediated by a space-time-modulated cryogenic-compatible Josephson metasurface through variation of spatiotemporal modulation parameters $\widetilde{\Phi}_\text{dc}$, $\widetilde{\Phi}_\text{rf}$, $d$, $f_\text{s}$, and $\gamma$. (a)~Selective excitation of Qubit 2 results in a state transition from its ground state to \( \hbar 2\omega_\text{s} \). (b)~Simultaneous excitation of Qubits 2 and 3 at distinct frequencies causes Qubit 2 to transition to \( \hbar 2\omega_\text{s} \) and Qubit 3 to transition to \( \hbar 3\omega_\text{s} \). (c)~Selective excitation of Qubit 3 leads to a state transition from its ground state to \( \hbar 3\omega_\text{s} \). (d)~Excitation of Qubits 3 and 4 at different frequencies drives Qubit 3 to \( \hbar 3\omega_\text{s} \) and Qubit 4 to \( \hbar 4\omega_\text{s} \). (e)~Selective excitation of Qubit 4 results in a state transition to \( \hbar 4\omega_\text{s} \). (f)~Excitation of Qubits 2, 3, and 4 at distinct frequencies enables transitions to \( \hbar 2\omega_\text{s} \), \( \hbar 3\omega_\text{s} \), and \( \hbar 4\omega_\text{s} \), respectively.}
		\label{Fig:res}
	\end{center}
\end{figure*}

\begin{figure*}
	\begin{center}
		\includegraphics[width=0.6\linewidth]{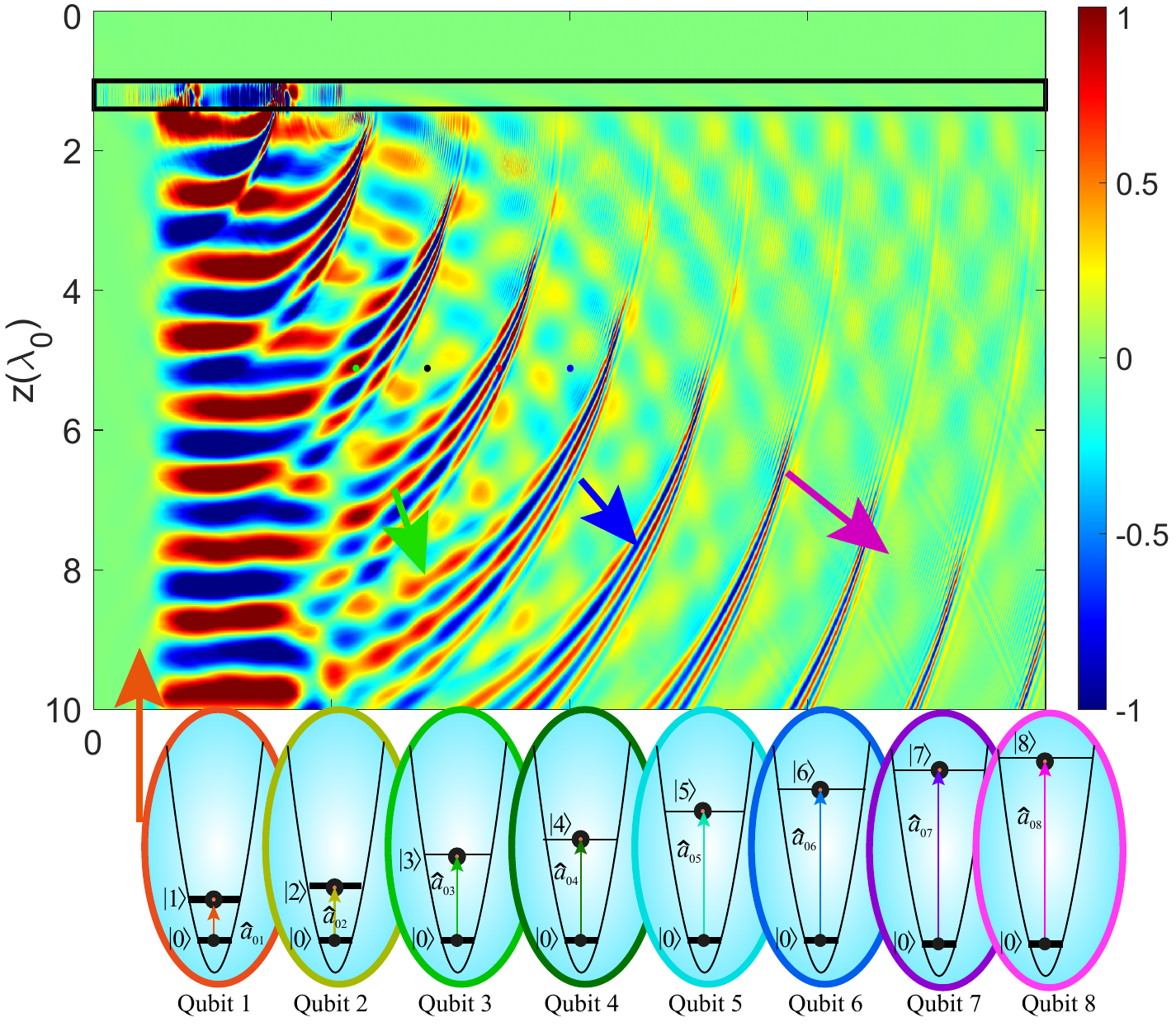}  
		\caption{Space-time coupling and quantum state transitions of eight superconducting qubits mediated by a space-time-modulated Josephson metasurface operating at \( f_\text{s} = 1.5 \, \text{GHz} \). The frequency detuning \( \Delta f = 1.5 \, \text{GHz} \) drives state transitions of Qubits 2 through 8 from their ground states to \( \hbar 2\omega_\text{s} \) through \( \hbar 8\omega_\text{s} \), respectively.}
		\label{Fig:res1p5}
	\end{center}
\end{figure*}

Figures~\ref{Fig:resa} through~\ref{Fig:resf} illustrate the selective space-time coupling and frequency-state transitions of superconducting qubits mediated by a space-time-modulated Josephson metasurface. These transitions are achieved through variations in the spatiotemporal modulation parameters, including $\widetilde{\Phi}_\text{dc}$, $\widetilde{\Phi}_\text{rf}$, metasurface thickness $d$, modulation frequency $\omega_\text{s}$, and the space-time velocity ratio $\gamma$. Figure~\ref{Fig:resa} shows the selective excitation of Qubit 2, which results in a state transition from its ground state to \( \hbar 2\omega_\text{s} \). The metasurface parameters used are $\widetilde{\Phi}_\text{dc}=1.22$, $\widetilde{\Phi}_\text{rf}=0.23$, $d=0.4 \lambda_0$, $f_\text{s}=3$ GHz, and $\gamma=1$. Figure~\ref{Fig:resb} depicts the simultaneous excitation of Qubits 2 and 3 at distinct frequencies, causing Qubit 2 to transition to \( \hbar 2\omega_\text{s} \) and Qubit 3 to \( \hbar 3\omega_\text{s} \). The metasurface parameters are $\widetilde{\Phi}_\text{dc}=1.2$, $\widetilde{\Phi}_\text{rf}=0.25$, $d=0.4 \lambda_0$, $f_\text{s}=3$ GHz, and $\gamma=1$. Figure~\ref{Fig:resc} highlights the selective excitation of Qubit 3, leading to a state transition from its ground state to \( \hbar 3\omega_\text{s} \). The metasurface parameters are $\widetilde{\Phi}_\text{dc}=1.2$, $\widetilde{\Phi}_\text{rf}=0.2$, $d=0.4 \lambda_0$, $f_\text{s}=6$ GHz, and $\gamma=1$. Figure~\ref{Fig:resd} shows the excitation of Qubits 3 and 4 at different frequencies, which drives Qubit 3 to \( \hbar 3\omega_\text{s} \) and Qubit 4 to \( \hbar 4\omega_\text{s} \). The metasurface parameters are $\widetilde{\Phi}_\text{dc}=0.97$, $\widetilde{\Phi}_\text{rf}=0.4$, $d=0.26 \lambda_0$, $f_\text{s}=3$ GHz, and $\gamma=0.6$. Figure~\ref{Fig:rese} demonstrates the selective excitation of Qubit 4, resulting in a state transition to \( \hbar 4\omega_\text{s} \). The metasurface parameters are $\widetilde{\Phi}_\text{dc}=0.95$, $\widetilde{\Phi}_\text{rf}=0.4$, $d=0.4 \lambda_0$, $f_\text{s}=3$ GHz, and $\gamma=0.5$. Figure~\ref{Fig:resf} presents the excitation of Qubits 2, 3, and 4 at distinct frequencies, enabling their respective transitions to \( \hbar 2\omega_\text{s} \), \( \hbar 3\omega_\text{s} \), and \( \hbar 4\omega_\text{s} \). The metasurface parameters are $\widetilde{\Phi}_\text{dc}=0.95$, $\widetilde{\Phi}_\text{rf}=0.4$, $d=0.4 \lambda_0$, $f_\text{s}=3$ GHz, and $\gamma=0.5$. This performance by the metasurface demonstrates its ability to achieve selective and simultaneous control over qubit transitions at distinct frequencies, enabling precise and scalable manipulation of quantum states for enhanced coherence, noise resilience, and entanglement fidelity in quantum computing systems.

Figure~\ref{Fig:res1p5} illustrates a scenario demonstrating the space-time coupling of eight qubits mediated by a space-time-modulated Josephson metasurface operating at a modulation frequency of \( f_\text{s} = 1.5 \, \text{GHz} \). In this configuration, the frequency detuning \( \Delta f = 1.5 \, \text{GHz} \) enables a sequential state transition of Qubits 2 through 8 from their ground states to \( \hbar 2\omega_\text{s} \) through \( \hbar 8\omega_\text{s} \), respectively. The metasurface's parameters are carefully optimized, with a DC modulation amplitude \( \widetilde{\Phi}_\text{dc} = 0.73 \), an RF modulation amplitude \( \widetilde{\Phi}_\text{rf} = 0.73 \), a metasurface thickness of \( d = 0.4 \lambda_0 \), and a space-time velocity ratio \( \gamma = 1 \). This performance underscores the metasurface's capability to achieve precise, multi-frequency coupling and quantum state transitions across a larger qubit array, paving the way for scalable and versatile quantum computing architectures. 

\section{Conclusions}\label{sec:conc} 
We introduced the concept of space-time-coupled qubits enabled by space-time-modulated cryogenic-compatible superconducting metasurfaces as a transformative platform to address key challenges in millikelvin-temperature quantum technologies. By leveraging the unique capabilities of space-time modulation, the proposed metasurface facilitates polychromatic coupling between qubits, enabling all-to-all connectivity, enhanced coherence, and robustness against noise and crosstalk. The metasurface's ability to mediate frequency-selective interactions provides significant advantages over traditional monochromatic architectures, including improved algorithmic efficiency, reduced gate operation errors, and extended coherence times. Our theoretical framework, supported by full-wave simulations and quantum performance analysis, demonstrates the effectiveness of this approach in enhancing the operational capacity of a 4 × 4 superconducting qubit array. The results indicate that frequency-division multiplexing, achieved through space-time modulation, not only minimizes interference but also stabilizes entanglement, offering a robust pathway for implementing quantum error correction and complex multi-qubit operations. The integration of space-time-modulated Josephson metasurfaces into quantum computing systems represents a paradigm shift in processor design, addressing critical limitations in current platforms and unlocking new avenues for scalable quantum architectures. Future research will explore the experimental realization of these metasurfaces and their integration with larger qubit arrays, paving the way for practical, high-performance quantum technologies capable of addressing the computational challenges of the future.

\newpage

\bibliography{Taravati_Reference.bib}

\end{document}